\documentstyle[prl,aps,twocolumn,psfig,floats]{revtex}

\begin{document}

\title{The liquid to vapor phase transition in excited nuclei}

\author{
J. B. Elliott, 
L. G. Moretto, 
L. Phair and 
G. J. Wozniak
}
\address{
Nuclear Science Division, Lawrence Berkeley National Laboratory, Berkeley, CA 94720\\
}

\author{
ISiS Collaboration\\
T. Lefort$^1$, 
L. Beaulieu$^2$, 
K. Kwiatkowski$^3$, 
W.-c. Hsi$^1$, 
L. Pienkowski$^4$, 
H. Breuer$^5$, 
R. G. Korteling$^6$,
R. Laforest$^7$, 
E. Martin$^7$,
 E. Ramakrishnan$^7$, 
D. Rowland$^7$, 
A. Ruangma$^7$,
V. E. Viola$^1$, 
E. Winchester$^7$ and 
S. J. Yennello$^7$
}
\address{
$^1$Department of Chemistry and IUCF, Indiana University, Bloomington, Indiana 47405\\
$^2$D\'{e}partment de Physique, Universit\'{e} Laval Qu\'{e}bec, Canada G1K 7P4\\
$^3$Los Alamos National Laboratory, Physics Division p-23, Los Alamos, NM 87545\\
$^4$Heavy Ion Laboratory, Warsaw University, Warsaw, Poland\\
$^5$University of Maryland, College Park, MD 20740\\
$^6$Department of Chemistry, Simon Fraser University, Burnaby, British Columbia, Canada V5A IS6\\
$^7$Department of Chemistry \& Cyclotron Laboratory, Texas A\&M University, College Station, TX 77843
}

\date{\today}
\maketitle

\begin{abstract}
For many years it has been speculated that excited nuclei would
undergo a liquid to vapor phase transition.\ \ For even longer, it has
been known that clusterization in a vapor carries direct information
on the liquid- vapor equilibrium according to Fisher's droplet
model.\ \ Now the thermal component of the $8$ GeV/c $\pi + ^{197}$Au
multifragmentation data of the ISiS Collaboration is shown to follow
the scaling predicted by Fisher's model, thus providing the strongest
evidence yet of the liquid to vapor phase transition.
\end{abstract}


\narrowtext

Nuclear multifragmentation, the break up of a nuclear system into
several intermediate sized pieces, has been frequently discussed in
terms of equilibrium statistical mechanics, and its possible
association with a phase transition
\cite{finn-82,siemens-83,moretto-97}.\ \ However, ample uncertainty
remains regarding its nature, in particular whether multifragmentation
is a phase transition and if so whether it is associated with the
liquid to vapor phase transition.

This paper will show that: 1) high quality experimental data contain
the unequivocal signature of a liquid to vapor phase transition
through their strict adherence to Fisher's droplet model; 2) the two
phase coexistence line is observed over a large energy/temperature
interval extending up to and including the critical point; 3) several
critical exponents, as well as the critical temperature, the surface
energy coefficient and the compressibility factor can be directly
extracted; 4) the nuclear phase diagram can be constructed with the
available data; 5) the nuclear liquid at break-up is a slightly
super-saturated vapor, for which the pressure and density can be
determined as a function of the temperature.

In past attempts to investigate the relationship between nuclear
multifragmentation and a liquid to vapor phase transition, critical
exponents have been determined
\cite{finn-82,gilkes-94,dagostino-99,elliott-00.1,elliott-00.2},
caloric curves have been examined \cite{pochodzalla-95}, and the
observation of negative heat capacities have been reported
\cite{dagostino-00}.\ \ Other studies of multifragmentation data have
shown two general, empirical properties of the fragment multiplicities
called reducibility and thermal scaling
\cite{moretto-97,beaulieu-98,moretto-99}.

Reducibility refers to the observation that for each energy bin the
fragment multiplicities are distributed according to a binomial or
Poissonian law.\ \ As such, their multiplicity distributions can be
{\it reduced} to a one-fragment production probability according to a
binomial or Poissonian distribution.

Thermal scaling refers to the feature that the average fragment yield
$\left< N \right>$ behaves with temperature $T$ as a Boltzmann factor:
$\left< N \right> \propto \exp( -B / T )$.\ \ Thus a plot of $\ln
\left< N \right>$ vs. $1 / T$, an Arrhenius plot, should be linear.\ \
The slope $B$ in such a plot is the one-fragment production
``barrier''.

Both the features of reducibility and thermal scaling are inherent to
any statistical model, in particular to the clusterization of droplets
from a vapor as described by Fisher's droplet model \cite{fisher-67}.\
\ Thus it is interesting to see if a system portraying reducibility
and thermal scaling portrays also the scaling of Fisher's model
\cite{elliott-00.2,mader-01}.\ \ Fisher's droplet model describes the
aggregation of molecules into clusters in a vapor, thus accounting for
its non-ideality.\ \ The abundance of a cluster of size $A$ is given
by:
	\begin{eqnarray}
	\label{fisher_droplet_a}
	n_A & = & q_0 A^{-\tau} \exp \left( \frac{A \Delta \mu}{T} +
	\frac{c_0 A^{\sigma}}{T_c} - \frac{c_0 A^{\sigma}}{T} \right)
	\\ \label{fisher_droplet_b} & = & q_0 A^{-\tau} \exp \left(
	\frac{A \Delta \mu}{T} - \frac{c_0 \varepsilon A^{\sigma} }{ T
	} \right) ,
	\end{eqnarray}
where $n_A$ is the number of droplets (or fragments) of mass $A$,
normalized to the size of the system $A_0$; $q_0$ is a normalization
constant depending only on the value of $\tau$ \cite{nakanishi-80};
$\tau$ is the topological critical exponent; ${\Delta}{\mu} = \mu -
{\mu}_l$, and $\mu$ and ${\mu}_l$ are the actual and liquid chemical
potentials respectively; $c_0 \varepsilon A^{\sigma}$ is the surface
free energy of a droplet of size $A$; $c_0$ is the zero temperature
surface energy coefficient; $\sigma$ is the critical exponent related
to the ratio of the dimensionality of the surface to that of the
volume; and $\varepsilon = ( T_c - T ) / T_c$ is the control
parameter, a measure of the distance from the critical point, $T_c$.

Equation~(\ref{fisher_droplet_a}) shows clearly that thermal scaling
is contained in this description.\ \ In particular, at coexistence
($\Delta \mu = 0$), Eq.~(\ref{fisher_droplet_a}) leads to a
``barrier'' $B = c_0 A^{\sigma}$.

	\begin{figure} [ht]
	\centerline{\psfig{file=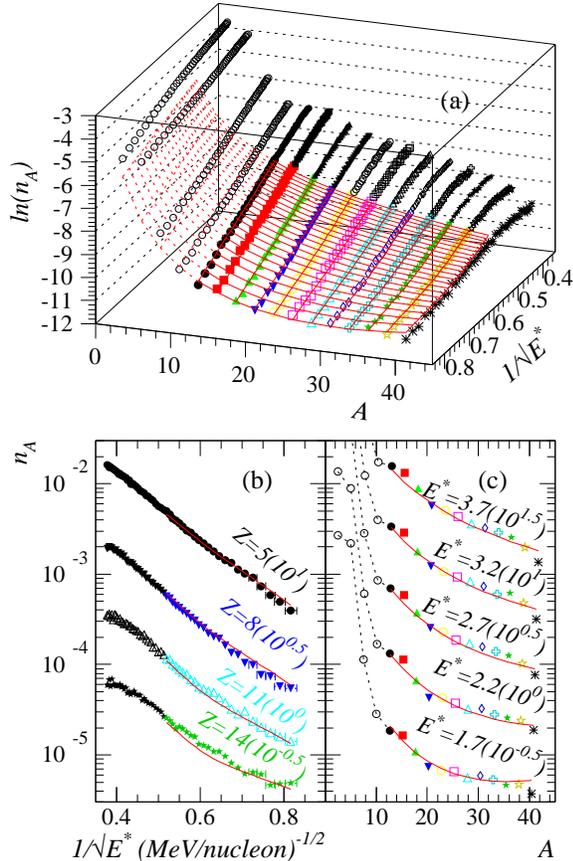,width=7.5cm,angle=0}}
	\caption{(a) The ISiS fragment yield surface: natural log of
	the fragment yield vs. fragment mass and inverse temperature.\
	\ (b) Arrhenius plots for representative charges.\ \ (c)
	Fragment mass yields for various values of $E^{*}$.\ \ Solid
	curves are from a fit to Fisher's droplet model; dashed curves
	show the extrapolation of the fit results.\ \ See
	Fig.~\ref{scaling} for symbol definition.\ \ Error bars
	(statistical) are smaller than the size of the points.}
	\label{three-d}
	\end{figure}

Recently, gold multifragmentation data from the ISiS Collaboration was
shown to exhibit reducibility and thermal scaling in the fragment
production probabilities \cite{beaulieu-00,beaulieu-01}.\ \ Since this
behavior is inherent to Fisher's model, it is interesting to determine
if Fisher's model describes the ISiS data set.\ \ In order to find if
this is the case, the ISiS charge yields from 8 GeV/c $\pi^- +$
$^{197}$Au fragmentation data (see Fig~\ref{three-d}a) were fit to
Eq.~(\ref{fisher_droplet_b}).

The mass of a fragment $A$ (estimated by multiplying the measured
fragment charge $Z$ by the $A$-to-$Z$ ratio of the fragmenting system)
was used as the cluster size in Eq.~(\ref{fisher_droplet_b}).\ \
The total number of fragments of a given size $N_A$ was normalized to
the size of the fragmenting system $A_0$ thus $n_A = N_A / A_0$.\ \
Here $\sqrt{E^*}$ was used in lieu of $T$, assuming the system
behaves as a degenerate Fermi gas.
 
The parameters of Fisher's model, e.g. $\tau$, $\sigma$, and
$E^{*}_{c}$ (in lieu of $T_c$) were used as fit parameters.\ \ The
distance from equilibrium $\Delta \mu$ was parameterized by a
polynomial of degree four in $E^*$ and the coefficients of that
polynomial were used as fit parameters.\ \ The surface energy
coefficient $c_0$ was parameterized by a polynomial of degree one and
the coefficients of that polynomial were used as fit parameters.\ \
The level density parameter was {\it absorbed} into the fit parameters
for $\Delta \mu$ and $c_0$.

While analyses similar to this one have been performed on
multifragmentation data in the past
\cite{hirsch-84,goodman-84,mahi-88}, those efforts dealt with
inclusive data sets; data from every excitation energy were examined
as a whole and the results were presented as a function of incident
beam energy.\ \ This work makes use of the high statistics, exclusive
data set of the ISiS Collaboration and bins the events in terms of
reconstructed excitation energy \cite{lefort-99}.\ \ In addition,
explicit use of Fisher's expressions for the bulk and surface
energies allows $\Delta \mu$ and $c_0$ to be determined by the data.

	\begin{figure} [ht]
	\centerline{\psfig{file=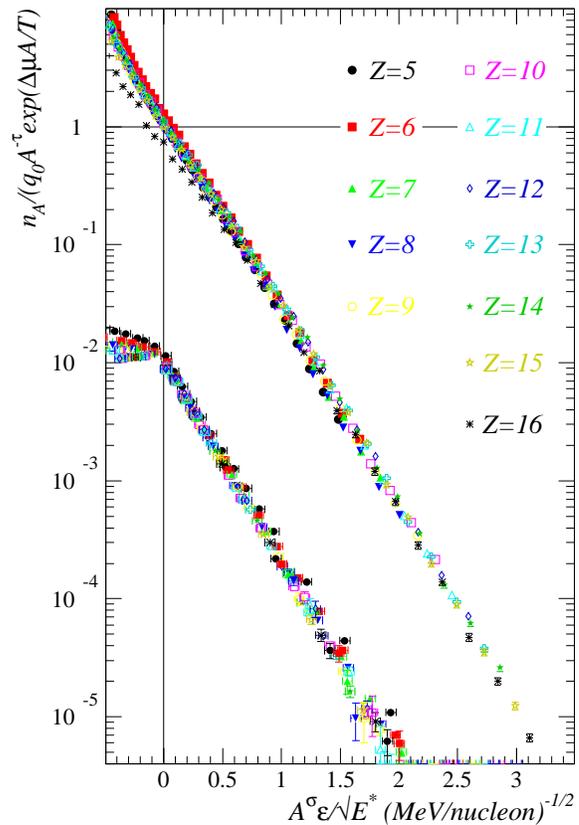,width=7.5cm,angle=0}} 
	\caption{The scaled fragment yield distribution versus the
	scaled temperature.\ \ ISiS data (upper) from a wide range in
	charge collapse to a single curve over more than five orders
	of magnitude.\ \ Also shown are similarly scaled data from a
	three-dimensional Ising model calculation (lower); the
	quantity $(n_A / q_0 A^{-\tau} ) / 100$ is plotted against the
	quantity $3.5(A^{\sigma} \varepsilon / T)$ for clusters of
	size $A = 5, 10, \ldots, 95, 100$.}
	\label{scaling}
	\end{figure}

Data for $E^* \le E^{*}_{c}$ and for $5 \le Z \le 16$ were included in
the fit to Eq.~(\ref{fisher_droplet_b}) and the parameters were
allowed to float to minimize ${\chi}^{2}_{\nu}$.\ \ Fisher's
parameterization of the surface energy $c_0 \varepsilon$ is invalid
for $T > T_c$, thus excitation energies greater than $E^{*}_{c}$ were
not considered.\ \ Fisher's model expresses the mass/energy of a
fragment in terms of bulk and surface energies.\ \ This approximation is
known to fail for the lightest of nuclei where structure details
(shell effects) dominate the mass.\ \ For this reason and the fact
that for the lightest fragments equilibrium and nonequilibrium
production cannot be clearly differentiated, fragments with $Z < 5$
were not considered in the fit.\ \ Fragments with $Z > 16$ were not
elementally resolved \cite{kwiatkowski-95}, and were also excluded.

	\begin{figure} [ht]
	\centerline{\psfig{file=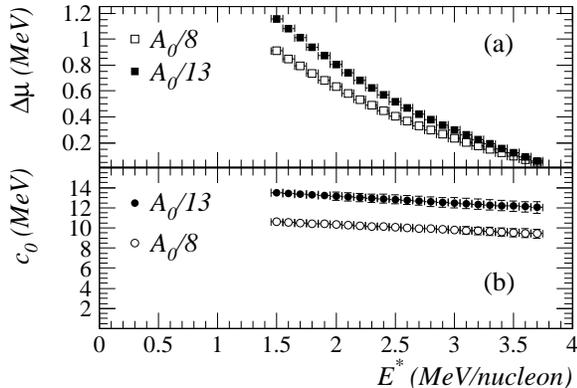,width=7.5cm,angle=0}}
	\caption{The behavior of $\Delta \mu$, and of the surface
	energy coefficient as a function of excitation energy up to
	$E^{*}_{c} = 3.7$ MeV/nucleon.\ \ See text for details.}
	\label{isis-dmu-c0-fig}
	\end{figure}

The behavior of the data for the $(n_A, A, T)$ surface is reproduced
over a wide range in excitation energy and fragment charge.\ \ This is
also shown in Fig.~\ref{three-d}.\ \ In this figure the results of the
analysis are presented in terms of Arrhenius plots
(Fig.~\ref{three-d}b), or of the fragment yield distribution plots
(Fig.~\ref{three-d}c).\ \ A powerful method to observe the results of
this analysis directly is to scale the data according to
Eq.~(\ref{fisher_droplet_b}) using the parameters resulting from the
fitting procedure.\ \ Figure~\ref{scaling} shows such a result.\ \ The
fragment mass yield distribution is scaled by the Fisher's power law
pre-factor and the bulk term: $n_A / q_0 A^{-\tau} \exp({\Delta}{\mu}
A /T)$.\ \ This quantity is then plotted against the temperature
scaled by Fisher's parameterization of the surface energy: $A^{\sigma}
\varepsilon / \sqrt{E^*}$.\ \ The scaled data collapse over five
orders of magnitude onto a single curve, which is precisely the
behavior predicted by Fisher's droplet model \cite{fisher-67}.\ \ This
curve is equivalent to a liquid-vapor coexistence line, as will be
shown below, and provides the best, most direct evidence yet for a
liquid to vapor phase transition in excited nuclei.

To illustrate the generality of this type of scaling,
Fig.~\ref{scaling} also shows the scaled cluster distributions from a
three-dimensional Ising model calculation \cite{mader-01}.\ \ This
system is know to model liquid-vapor coexistence up to the critical
temperature.\ \ The perfect scaling of the cluster yields according to
Eq.~(\ref{fisher_droplet_b}) demonstrates liquid-vapor-like
coexistence up to $T_c$.

	\begin{figure} [ht]
	\centerline{\psfig{file=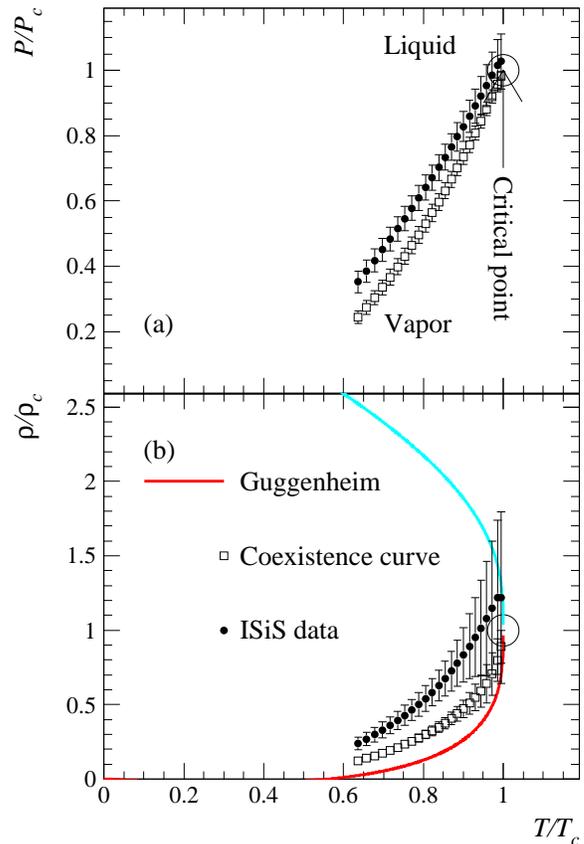,width=7.5cm,angle=0}}
	\caption{(a) The reduced pressure-temperature plot and (b) the
	reduced density-temperature plot.\ \ Filled symbols show the
	path of the ISiS data, open symbols show the coexistence
	curve of charged finite nuclear matter.}
	\label{temp-den}
	\end{figure}

The value of $\tau$ extracted from this analysis, $\tau = 2.00 \pm
0.01$, is in the range predicted by Fisher's model.\ \ Similarly,
$\sigma = 0.64 \pm 0.05$ is close to the value expected for a three
dimensional system, $\sim 2/3$.\ \ The values of $\Delta \mu$ and
$c_0$ and their energy dependence are shown in
Fig.~\ref{isis-dmu-c0-fig}.\ \ The positive value of $\Delta \mu$ for
$E^* < E^{*}_{c}$ indicates that the system behaves as a slightly
super-saturated vapor.\ \ Values for $\Delta \mu$ and $c_0$ can be
determined by making a choice of the level density parameter.\ \ For
$a = A_0 / 8$, $c_0 \approx 10.0 \pm 0.3$ MeV while $a = A_0 /
13$ gives $c_0 \approx 12.8 \pm 0.4$ MeV; both values are in the
range of the value of the surface energy coefficient of the
semi-empirical mass formula ($\sim 17$ MeV), although the linear
extrapolation all the way down to $T = 0$ implied by
Eq.~(\ref{fisher_droplet_b}) is questionable.\ \ The values of the
critical exponents determined here are in agreement with those
determined for the EOS multifragmentation data \cite{elliott-00.1} and
the value of the excitation energy at the critical point $E^{*}_{c} =
3.70 \pm 0.05$ MeV/nucleon is in the neighborhood of the value
observed in the EOS analysis ($E^{*}_{c} = 4.75$ MeV/nucleon)
\cite{elliott-00.1,elliott-00.2,hauger-98}.\ \ The EOS analysis relied
on the assumption that $\Delta \mu \approx 0$.\ \ No such assumption
was made in this work.\ \ In fact the present effort tests that
assumption and finds it to be approximately valid below the critical
point since the observed $\Delta \mu$ values are at most $\sim 1$ MeV
(assuming $a = A_0 / 13$).

Using the value of $E^{*}_{c}$ determined here and a level density
parameter of $A_0 / 13$, the critical temperature can be estimated to
be on the order of $6.9$ MeV, which is comparable to theoretical
estimates for small nuclear systems \cite{jaqaman-84,bonche-85}.\ \ In
most theoretical calculations the Coulomb force and small size of the
system drastically reduces the value of $T_c$.\ \ It is also well
known that both $T_c$ and ${\rho}_c$ scale as a function of system
size in many systems \cite{wilding-95,ma-00}.

The coexistence line and actual position of the fragmenting system in
the pressure-density-temperature $(P, \rho, T)$ diagram can be
determined from this analysis.

Fisher's theory assumes that the non-ideal vapor can be approximated
by an ideal gas of clusters.\ \ Accordingly, the quantity $n_A$ is
proportional to the partial pressure of a fragment of mass $A$ and the
total pressure due to all of the fragments is the sum of their partial
pressures:
	\begin{equation}
	\frac{P}{T} = \sum_{A=1}^{\infty} n_A .
	\label{pressure}
	\end{equation}
The reduced pressure is then given by:
	\begin{equation}
	\frac{P}{P_c} = \frac{\sum_{A=1}^{\infty}
	n_A(T)}{\sum_{A=1}^{\infty} n_A(T_c)}.  
	\label{reduced_pressure}
	\end{equation}
This is a transformation of the information given in
Fig.~\ref{isis-dmu-c0-fig} onto a more familiar frame of reference.

When values of $n_A$ corresponding to $\Delta \mu = 0$ are used in
these equations, the coexistence curve is obtained.\ \ In other words,
this analysis provides simultaneously the coexistence line and the
actual line.\ \ Both are shown in Fig~\ref{temp-den}a.

The system's density can be found via
	\begin{equation}
	\rho = \sum_{A=1}^{\infty} A n_A ,
	\label{density}
	\end{equation}
and the reduced density from
	\begin{equation}
	\frac{\rho}{{\rho}_c} = \frac{\sum_{A=1}^{\infty}
	A n_A(T)}{\sum_{A=1}^{\infty} A n_A(T_c)} .
	\label{reduced_density}
	\end{equation}
As before, both the actual curve and the coexistence curve can be
determined.\ \ Both are shown in Fig.~\ref{temp-den}b together with
the coexistence curve of Guggenheim \cite{guggenheim-45}.

The coexistence curve from the nuclear data does not agree closely with
the Guggenheim plot over the full range of $T / T_c$.\ \ This is not
surprising given the complexity of the nuclear fluid compared to the
simple fluids analyzed by Guggenheim.\ \ However, near $T_c$, the
universal behavior of the coexistence curve is recovered.

The compressibility factor $p_c / T_c {\rho}_c$ is calculated by
dividing the denominator of Eq.~(\ref{reduced_pressure}) by that of
Eq.~(\ref{reduced_density}) and found to be $0.24 \pm 0.06$.\ \ This
value agrees with the value of the compressibility factor of many
fluids \cite{kiang-70}.

In conclusion this paper has shown that the data of the ISiS
Collaboration contain the unequivocal signature of a liquid to vapor
phase transition via their strict adherence to Fisher's droplet
model.\ \ Through Fisher's scaling of the fragment yield distribution
(Fig.~\ref{scaling}) the two phase coexistence line has been
determined over a large energy/temperature interval extending up to
and including the critical point.\ \ The critical exponents $\tau$ and
$\sigma$ as well as the approximate value of $T_c$, the surface energy
coefficient $c_0$ and the compressibility factor have been extracted
and agree with accepted values.\ \ A portion of the nuclear phase
diagram has been constructed with the available data.\ \ Finally, the
nuclear liquid at break-up was observed to be a slightly
super-saturated vapor for which $P / P_c$ and $\rho / {\rho}_c$ were
determined as a function of $T / T_c$.

The authors would like to thank Prof. Cathy Mader for her input and
invaluable efforts with the Ising model calculations.\ \ This work was
supported by the the US Department of Energy, National Science
Foundation, the National Science and Engineering Research Council of
Canada, the Polish State Committe for Scientific Reseach, Indiana
Universtiy Office of Research, the University Graduate School, Simon
Fraser University and the Robert A. Welch Foundation.

\end{document}